\documentclass[10pt]{article}
\usepackage[OE]{express}

\begin{document}

\title{Time-resolved high harmonic spectroscopy of
dynamical symmetry breaking in bi-circular laser fields}

\author{\'Alvaro Jim\'enez-Gal\'an,\authormark{1,*} Nickolai Zhavoronkov,\authormark{1} Marcel Schloz,\authormark{1} Felipe Morales,\authormark{1} and Misha Ivanov \authormark{1,2,3}}

\address{\authormark{1}Max-Born Institute for Nonlinear Optics and Short Pulse Spectroscopy, Max-Born-Stra{\ss}e 2A, D-12489 Berlin, Germany\\
\authormark{2}Institute f\"ur Physik, Humboldt-Universit\"at zu Berlin, Newtonstra{\ss}e 15, D-12489 Berlin, Germany\\
\authormark{3}Department of Physics, Imperial College London, South Kensington Campus, SW72AZ London, UK}

\email{\authormark{*}jimenez@mbi-berlin.de} 



\begin{abstract}
The bi-circular scheme for high harmonic generation, which combines two counter-rotating circular fields with frequency ratio 2:1, has recently permitted to generate
high harmonics with essentially circular polarization, opening the way for ultrafast chiral studies. This scheme produces harmonic lines at $3N+1$ and $3N+2$ multiples of the fundamental driving frequency, while
the $3N$ lines are forbidden owing to the three-fold symmetry of the field. It is generally established that the routinely observed signals at these forbidden harmonic lines come from a slight ellipticity in the driving fields,
which breaks the three-fold symmetry. We find that this is neither the only nor it is the dominant mechanism responsible. The forbidden lines can be observed even for perfectly circular, long driving pulses. We show that they encode rich information on the sub-cycle electronic dynamics that occur during the generation process. By varying the time delay and relative intensity between the two drivers, we demonstrate that when the second harmonic either precedes or is more intense than the fundamental field, the dynamical symmetry of the system is broken by electrons trapped in Rydberg orbits (i.e., Freeman resonances), and that the forbidden harmonic lines are a witness of this.
\end{abstract}





\section{Introduction}
High harmonic generation serves as an indispensable source of bright, coherent XUV and soft X-ray light. This
light is used
to induce, monitor and control the dynamics of electrons in atoms, molecules and solids at their intrinsic timescale \cite{Krausz2009, Sansone2010, Gruson2016, Kelkensberg2011, Hassan2016, Garg2016}.
High harmonic generation  also serves
as a spectroscopic tool for unraveling the complex multi-electron and coupled electron-nuclear dynamics in
molecules \cite{baker2006probing,smirnova2009high, haessler2010attosecond, vozzi2011generalized, shafir2012resolving, worner2010following, ferre2016two, Morales2012, pedatzur2015attosecond, bruner2015multidimensional, bruner2016multidimensional, serbinenko2013multidimensional, Smirnova2015, kraus2015measurement, worner2011conical}
 and solids  \cite{ghimire2011observation, schiffrin2013optical, ivanov2013opportunities, vampa2014theoretical, vampa2015linking, langer2016lightwave}.

From the light source perspective, high harmonic generation in two counter-rotating circularly polarized fields
made by superposing the fundamental field $\omega$
and its second harmonic  $2\omega$ \cite{Eichmann1995,Milosevic2000} is particularly appealing
\cite{fleischer2014spin,Chen2016, Hickstein2015, Kfir2015,Baykusheva2016,bandrauk2016circularly,mauger2016circularly,odvzak2016atomic,
odvzak2016high,Fan2015}.
It allows one to generate XUV radiation and trains of attosecond pulses with controlled polarization properties
\cite{Milosevic2000, fleischer2014spin, Milosevic2015} and carries the potential to generate isolated circularly polarized attosecond pulses
\cite{Medisauskas2015, Zhang2017}. These pulses would open the route
to studying chiral-sensitive light-matter interactions with unprecedented temporal resolution in
gas and condensed phase, e.g., the study of ultrafast chiral-specific dynamics in molecules, ultrafast chiral recognition via photoelectron circular dichroism \cite{Cireasa2015, Smirnova2015}, ultrafast magnetization and spin dynamics \cite{Cavalieri2007, Boeglin2010, Graves2013}, etc.

While the potential of high harmonic generation in bi-circular fields as a light source is actively explored,
the complementary {\it spectroscopic} potential of this  scheme for studying the underlying
electronic dynamics and dynamical symmetries  is far less known.  Pertinent recent papers include Refs.
\cite{Smirnova2015,Baykusheva2016,mauger2016circularly,odvzak2016high}.
This situation stands in stark contrast to the two-dimensional high harmonic spectroscopy which
uses the combination of linearly polarized fundamental and its second harmonic
\cite{dudovich2006measuring, mansten2008spectral, mauritsson2009sub, he2010interference, dahlstrom2011quantum,
brugnera2011trajectory, ganeev2012experimental, shafir2012resolving, serbinenko2013multidimensional, ferre2016two, Morales2012, pedatzur2015attosecond, bruner2015multidimensional, bruner2016multidimensional},
allowing one to track electronic and vibronic
\cite{ferre2016two} dynamics with temporal resolution from tens of femtoseconds down to
tens of attoseconds.

Here we demonstrate the spectroscopic potential of high harmonic generation
in bi-circular laser fields to track light-driven dynamical symmetry breaking in a quantum
system. In general, the emergence of strong, symmetry forbidden, lines in high harmonic spectra
is a tell-tale sign of symmetry breaking induced by the underlying attosecond electronic
\cite{ivanov1993coherent, smirnova2007anatomy} or vibronic dynamics
\cite{morales2014high, bian2014probing, Silva2016}. Specifically,
we show that symmetry forbidden lines in high harmonic spectra generated in
bi-circular fields are sensitive to frustrated tunnel ionization
\cite{yudin2001physics, nubbemeyer2008strong, eichmann2009acceleration, von2013frustrated, eichmann2013observing, zimmermann2017unified} and the presence of strongly laser-driven Rydberg states,
the so-called `bound states of the free electron'   \cite{richter2013role}, which are able to survive
intense laser fields \cite{popov2003strong, popov2011different, fedorov2012interference, nubbemeyer2008strong, eichmann2009acceleration, morales2011imaging, von2013frustrated, eichmann2013observing, zimmermann2017unified}
even when the ground state of the neutral is completely depleted
\cite{eichmann2009acceleration, bredtmann2016xuv}.


In contrast to single-color high harmonic spectroscopy of the dynamical symmetry breaking
\cite{ivanov1993coherent, morales2014high, bian2014probing, Silva2016},
the two-color laser field offers clear advantages: it allows one to tune the time-delay
between the two colors and their relative intensities. We rely on this ability in
the present work. It allows us to make
first steps towards adressing an extremely exciting but equally challenging
problem of time-resolving the frustrated tunneling process \cite{Yudin2001,
nubbemeyer2008strong, eichmann2009acceleration, manschwetus2009strong, eichmann2013observing}
during the driving laser pulse.

The ability to control the shape of the driving field
by changing the relative  intensities of the two colors and
their delay also brings up the
complementary aspect of attosecond electron dynamics in multi-color fields --
the ability to control these dynamics and the properties of the emitted radiation
\cite{dudovich2006measuring, mansten2008spectral, mauritsson2009sub,
brugnera2011trajectory, Morales2012}. We explore this ability in
the present work.

\begin{figure}[hbpt!]
\centering
\includegraphics[width=\linewidth]{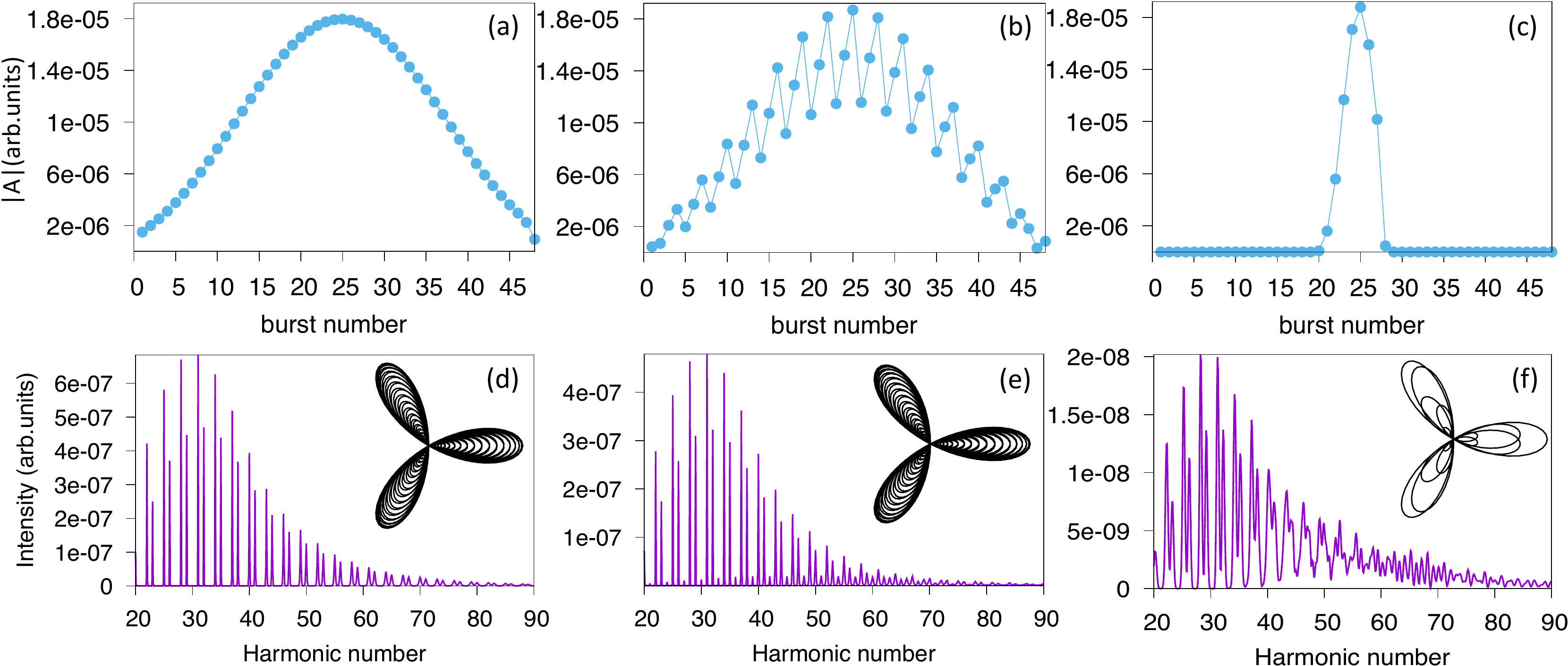}
\caption{\label{fig:sketch} {\bf Typical harmonic spectra in bi-circular fields.}
Strong field approximation solutions for the bicircular scheme using a short-range potential with an ionization potential of $I_p=24.6$~eV (as helium).
In the top row, the amplitude distribution of the different bursts, temporally ordered, that contribute to an arbitrary high harmonic (H60).
In the bottom row, the HHG spectra. The fields are Gaussian-shaped, with a peak intensity of $I=3.5\times 10^{14}$ W/cm$^2$, and duration
of: (a,d) 38~fs, (b,e) 38~fs with an ellipticity of 0.9, and (c,f) 6~fs. The Lissajous figures of the corresponding fields are shown in the top right corners of the bottom panels.}
\end{figure}

When a circularly polarized driver with
frequency $\omega$ is used in combination with its counter-rotating second harmonic,
the resulting field  has the three-fold symmetry shown in Fig.\ref{fig:sketch}. As a consequence,
high harmonic spectra generated in centrally symmetric media present
peaks at the $3N+1$ and $3N+2$ harmonic lines, but not at $3N$. The $3N+1$ and $3N+2$ harmonics
are circularly polarized and rotate in the directions of the $\omega$ and $2\omega$ fields, respectively.
The $3N$ harmonic lines are symmetry forbidden, their lack reflecting the conservation of the
angular momentum. Indeed, these lines correspond to the absorption of
the net amount $N$ of the fundamental $\hbar\omega$ photons
and the net amount $N$ of the second harmonic $2\hbar\omega$ photons, i.e.
the net total of $2N$ photons, preserving the partity of the initial state and thus
precluding one-photon radiative recombination to it.

In spite of this clear symmetry argument,  non-negligible signals at
3$N$ harmonic lines have been routinely observed in experiments,
starting with the pioneering work  \cite{Eichmann1995}.
Their presence has been systematically ascribed to
slight ellipticity of the drivers. While this is certainly an important experimental reason, it is
not the only one, as has been recently highlighted by Baykusheva \cite{Baykusheva2016}.
The emergence of strong forbidden lines can manifest the lack of
symmetry of the quantum system, in particular the destruction of the dynamical
symmetry within the laser cycle (Fig.\ref{fig:sketch}).
This makes the analysis of the forbidden lines, ideally in a time-resolved fashion,
very interesting, opening a route to time-resolving the changing symmetries
of the quantum system.

Turning from the spectroscopic aspect of high harmonic
generation to the light-source aspect, it is also important to understand the
origin of the forbidden lines, the mechanisms controlling their strengths and
polarization. Indeed, these lines will play crucial role in determining the
polarization properties of attosecond pulses or pulse trains produced
by the combination of circularly polarized
high harmonics generated in a bi-circular field.

Addressing these issues  is the focus of this paper.
In particular, we find theoretically that small deviations from perfectly circular light $\epsilon=1.0$, e.g.
$\epsilon =0.95$, which would be typical for realistic experiments,
is hardly the main reason for their prominence.
Thus, the emergence of strong forbidden lines in standard experiments with nearly circular
pulses is rather unexpected and cannot be blamed entirely on small deviations from
perfect circularity.

To uncover the physics responsible for  the $3N$ lines, we study the case when the
two pulses constituting the bicircular field, $\omega$ and $2\omega$, are time-delayed
but still overlap. This allows us to track the emergence of the $3N$
lines as a function of the  $\omega - 2\omega$ delay. Using  Helium as a target gas and
the combination of 800 nm and 400 nm driving fields,
we find experimentally and theoretically that the $3N$ lines become stronger as the delay between the
two pulses increases and their overlap decreases, especially when
the $2\omega$ (400 nm) pulse comes first.

It is well known that, in contrast to
800 nm, the 400 nm pump leads to efficient accumulation of
population in Rydberg states via frustrated tunnelling
(see e.g. \cite{zimmermann2017unified} for detailed expermental and theoretical analysis),
and that these states survive strong dressing fields
\cite{Yudin2001,
nubbemeyer2008strong,eichmann2009acceleration,manschwetus2009strong,eichmann2013observing}.
Thus, our results suggest that in the pump-probe type setup, when
the 800 nm pulse is delayed, the 400 nm field excites the  bound states
and  breaks the dynamical symmetry due to sub-cycle accumulation of population in these states.
Since frustrated tunnelling in the 800 nm field is less
efficient than for 400 nm field \cite{zimmermann2017unified},
the dynamical symmetry breaking should be weaker when the 800 nm pulse
comes first. This expectation is confirmed by our observations: the forbidden
3$N$ lines are more prominent when the 400 nm pulse precedes the 800 nm
pulse.

The importance of Rydberg excitation is further tested experimentally by changing the intensity
of the 400 nm pulse. In agreement with the above physical picture, we find
that the forbidden lines become more prominent with higher intensity of the 400 nm
light.
Theoretically, even for perfectly circular pulses, the forbidden harmonic lines appear
in a dramatic way if the intensity of the 400 nm field is increased substantially  above
the fundamental. In contrast, raising the 800 nm intensity does not have the same effect.

We further confirm this physical picture by showing how the gradual build up of the forbidden lines in the spectrum.
Thus,  from the high harmonic spectroscopy perspective, analyzing the appearance of
the forbidden harmonics as a function
of the $\omega$-$2\omega$ delay and intensities, we make first steps towards seeing
how the frustrated tunneling process \cite{Yudin2001,
nubbemeyer2008strong,eichmann2009acceleration,manschwetus2009strong,eichmann2013observing}
unfolds in time.

From the light source perspective, we analyze the unusual polarization
properties of the 3$N$ lines and show the ways of controlling their
strength and ellipticity: by varying the
two-color delay or the relative intensities of the two driving fields.
The intensities and the polarization properties of these lines
are important in determining the
polarization properties of the attosecond pulse trains produced
via high harmonic generation in bi-circular fields.

\subsection*{Dynamical symmetry and the selection rules}

Consider high harmonics generated by the two counter-rotating circular pulses with
frequencies $\omega$ and 2$\omega$. We write the total electric field as
\begin{equation}\label{eq:electric_field}
\mathbf{E}(t) = \sqrt{2}\,\mathrm{Re} \left\{ -F_{\omega} e^{-i\omega t} \hat{\mathbf{e}}_+ + F_{2\omega} e^{-i 2\omega t} \hat{\mathbf{e}}_- +0\,\hat{\mathbf{e}}_0\right\}.
\end{equation}
The three components of the field correspond to the
counter-clockwise ($\hat{\mathbf{e}}_+ = -(\hat{\mathbf{e}}_x + i\hat{\mathbf{e}}_y)/\sqrt{2}$) and clockwise ($\hat{\mathbf{e}}_- = (\hat{\mathbf{e}}_x - i\hat{\mathbf{e}}_y)/\sqrt{2}$) rotations in the $x$-$y$ plane,
and a linear component ($\hat{\mathbf{e}}_0 = \hat{\mathbf{e}}_z$) along the $z$ axis. For collinear driving fields,
which we assume in this work, the latter is always zero. The fundamental field rotates in the
counter-clockwise, positive direction (polarization $\hat{\mathbf{e}}_+$), while the second harmonic rotates
clockwise (polarization $\hat{\mathbf{e}}_-$).

The inset in Fig~\ref{fig:sketch}d shows the total field, which has characteristic
three-leaf structure. The total field rotates counter-clockwise,
from the leaf $k=1$ aligned horizontally, along the x-axis,
to the leaf $k=2$ turned 120 degrees counter-clockwise, to the leaf $k=3$ turned  240 degrees counter-clockwise.
Each leaf generates a burst of emission.
To simplify the discussion and notations,  but without the loss of generality,
we consider emission associated with the so-called short trajectories
\cite{agostini2004physics}. In bicircular fields, their contribution on the single-atom level is
dominant  \cite{Milosevic2000}.
Within each cycle of the fundamental field, the emission contains three
bursts, each associated with one of the three leafs of the total field and labelled by the
index $k=1,2,3 $. The corresponding induced dipoles are
\begin{equation}\label{eq:DipoleSum}
\mathbf{d}^{(k)}(t)=  \mathbf{A}_+^{(k)}(t)\,e^{i\mathbf{\phi}_+^{(k)}(t)}\hat{\mathbf{e}}_+ +
\mathbf{A}_-^{(k)}(t)\,e^{i\mathbf{\phi}_-^{(k)}(t)}\hat{\mathbf{e}}_-   .
\end{equation}
Each dipole has components with both polarizations, $\hat{\mathbf{e}}_{\pm}$, and each
component is carrying its own amplitude and phase. As always in strong-field driven
high harmonic generation, the phase is dominated
by the action accumulated during the motion of the electron in the continuum, while
the amplitude is dominated by strong-field ionization conditioned on the
electron return to the vicinity of the parent ion. Note that the two-color
driving field curves the electron trajectory. This
curved continuum motion imposed by the field
means that the recombination dipole matrix elements will
be different for the emission of photons co-rotating and counter-rotating with the
fundamental laser field.

If the circular pulses at the frequencies $\omega$ and $2\omega$ are long and
overlap perfectly, then a rotation
of $2\pi k/3$ (with $k$ integer) leaves the field invariant (see Fig.~\ref{fig:sketch}).
This three-fold symmetry of the field, together with
the symmetry of the medium, imply that
the phases and the amplitudes for $k=2,3$ are the same as for $k=1$,
up to the $2\pi/3$ and $4\pi/3$ rotations  of the associated
vectors $\hat{\mathbf{e}}_{\pm}$ for $k=2$ and $k=3$,
and the time-delay of the emission bursts by $1/3$ and $2/3$ of the laser cycle correspondingly.
Under  rotation by an angle $\alpha$, the $\hat{\mathbf{e}}_{\pm}$ vectors  transform as
\begin{eqnarray}\label{eq:Rotation}
\hat{\mathbf{e}}_+  \rightarrow \hat{\mathbf{e}}_+ e^{-i\alpha}
\nonumber
\\
\hat{\mathbf{e}}_- \rightarrow \hat{\mathbf{e}}_- e^{+i\alpha}
  .
\end{eqnarray}
Upon the Fourier transform into the frequency domain, the contributions from each leaf
will gain additional phases $\exp(iMk 2\pi/3)$ from the $\exp(iM\omega t)$ factor in the Fourier
integral, due to the
corresponding time-delays in emission by $\omega t=k 2\pi/3$. Hence, the total contribution of the
three bursts to the emission dipole at the frequency $M\omega$ is, for the
component co-rotating with the fundamental field,
\begin{eqnarray}
\label{eq:FT1}
&&\mathbf{d}_{+}(M\omega)=\mathbf{A}_+(M) e^{i\mathbf{\phi}_+(M)}\hat{\mathbf{e}}_+
\left[
1+e^{-i2\pi/3+iM 2\pi/3}+e^{-i4\pi/3+iM 4\pi/3}
\right]
\end{eqnarray}
Similarly, for the component of the harmonics which co-rotates with the second harmonic, we
have
\begin{eqnarray}
\label{eq:FT2}
&&\mathbf{d}_{-}(M\omega)=\mathbf{A}_-(M) e^{i\mathbf{\phi}_-(M)}\hat{\mathbf{e}}_-
\left[
1+e^{+i2\pi/3+iM 2\pi/3}+e^{+i4\pi/3+iM 4\pi/3}
\right]
\end{eqnarray}
For the intensities  $I_{\hat{\mathbf{e}}_{\pm}}\propto \left|\mathbf{d}_{\pm}(M\omega)\right|^2$
we obtain
\begin{eqnarray}\label{eq:IntensityModulation}
&&
I_{\hat{\mathbf{e}}_+}(M\omega)\propto 3 + 2\cos \left[2\pi (M -1)/3\right] + \cos\left[4\pi (M-1)/3\right],
\nonumber
\\
&&
I_{\hat{\mathbf{e}}_-}(M\omega)\propto 3 + 2\cos \left[2\pi(M+1)/3\right] + \cos\left[4\pi (M+1)/3\right].
\end{eqnarray}

The expressions Eqs.(\ref{eq:FT1},\ref{eq:FT2},\ref{eq:IntensityModulation})
show the symmetry-imposed selection rules for the harmonics of different
polarization. We stress again that  these rules assume that three emission bursts
associated with the  three leaves of the driving field are identical, up to rotation and time-delay.

The counter-clockwise component, co-rotating with the red field, will be enhanced
for the $M=3N+1$ harmonics and cancelled by the interference of the
three terms for the $M=3N+2$ harmonics.
The clockwise component, co-rotating with the blue field, will be
enhanced the $M=3N+2$ harmonics and cancelled by the
interference of the three terms for $M=3N+1$ harmonic.
Finally, the $M=3N$ harmonics will be always suppressed
(provided the fields are propagating collinearly), giving rise to the characteristic HHG spectrum
of the $\omega+2\omega$ scheme \cite{Eichmann1995,Milosevic2000}(Fig.~\ref{fig:sketch}d).
Fig.~\ref{fig:sketch}(a) shows the calculated amplitudes $|A|$ of the different emission bursts
for a given harmonic (we have used $M=60$ in this Figure)
using an ellipticity of 1.0 in both fields, in the case when the two driving fields
have identical pulse durations and overlap perfectly. The calculations are based on using
the saddle point method and the standard strong field approximation (SFA),  following
Milosevic and Becker \cite{Milosevic2000}
and considering the contribution of the saddle points associated with the
short trajectories in the bicircular field.  The ionization potential was set to
$I_p=24.6$~eV (Helium), with the ground  s-state being the initial state.
The driving pulses are both Gaussian-shaped, with a peak field
strength of $F_{\omega}=F_{2\omega}=0.1$ a.u., and duration of
(a,d) 38~fs. The harmonic spectrum, shown in Fig.~\ref{fig:sketch}(d),
demonstrates the lack of M=60 and all harmonics with orders $M=3N$.
Interestingly, the lines $M=3N+1$ and $3N+2$, with opposite helicity, have different heights,
even though we used the ground s-state. This propensity reflects the curvature of the
motion imposed on the electron between ionization and recombination. Here,  the
curvature is dominated by the fundamental field.

\subsection*{Physical origin of the forbidden harmonics}

Several factors can alter the simple selection rules associated with 
the three-fold dynamical symmetry of the driving field. 
Two possibilities lie on the surface and are illustrated
in Fig.~\ref{fig:sketch}(b,c,e,f).

First, if the pulses are elliptical rather than perfectly circular,
the field will not be invariant under a $2\pi/3$ rotation anymore. 
Fig.~\ref{fig:sketch}b shows the amplitudes $|A|$
for a given harmonic (we have used $M=60$ in this Figure)
the different emission bursts, but now
for the ellipticity of 0.9 (for both fields).
To eliminate other possible mechanisms responsible for the 3$N$ lines
and focus on the role of ellipticity, the calculations used the strong field approximation (SFA)
approach for the bi-circular fields \cite{Milosevic2000}.
One of the bursts inside the cycle is stronger than the other two, leading to the appearance of the
forbidden harmonics in the spectrum (see Fig.~\ref{fig:sketch}(e)
This assymmetry in the driving field is hardly visible in Fig.~\ref{fig:sketch}(e) inset,
but becomes noticeable in the harmonic amplitudes due to exponential sensitivity of tunnel ionization
to the field. Nevertheless, the forbidden lines are quite weak, even for
these rather substantial deviations from the perfectly circular driving fields.

Second, for short pulses, many of the bursts will be strongly suppressed by a rapidly changing
envelope. Their amplitudes and phases, for a specific harmonic,
will depend on the rapidly changing fields at the times of ionization and recombination. In particular,
this leads to a heavily non-symmetric amplitude distribution with respect to the central
(more intense) burst, as can be seen in Fig.~\ref{fig:sketch}(c), again for H60. This loss of symmetry can be
observed in the HHG as a signal at the forbidden harmonic lines. These indeed arise
prominently for H60 and higher orders (Fig.~\ref{fig:sketch}(f).

There are, however, two additional, more subtle, possibilities of breaking the
dynamical symmetry that are crucial in this work.
First, a memory present in the quantum system
would make the contributions from successive peaks of the field different.
This is in analogy to XUV-assisted high harmonic generation in linear fields \cite{Ishikawa2003}, where
an XUV pulse pumps the system to a superposition of Rydberg states, from which ionization
occurs easier. In this case, the $400$~nm pulse acts as a multi-photon pump which excites the system into
a superposition of Rydberg states. Ionization in the subsequent bursts is therefore enhanced and hence
$A_{+/-}^{(1)} \neq A_{+/-}^{(2)} \neq A_{+/-}^{(3)}$, which breaks the
dynamical symmetry.
Second,  one can
vary the time delay between the two driving pulses, the
fundamental and the second harmonic, breaking the
symmetry of the total field in a well controlled way.
For the two circular driving fields,
high harmonic signal will only be produced in the region of their overlap. Time delaying
one of the two pulses  leads to the asymmetric behaviour of the successive
emission bursts, similar to a short two-color pulse. We use the interplay of these two possibilities
in our analysis below.

The physical idea is as follows:
the system memory is linked to the excitations generated by the driving
pulse. The efficiency of the excitations by the second harmonic field
is higher than by the fundamental. Hence, memory effects
and the strengths of the forbidden harmonics should be
more prominent when the second harmonic arrives first, compared to
the case when the second harmonic arrives second.  Thus,
systematically varying the  two-color delay and recording the relative strength
of the forbidden lines allows us to gauge the role of the dynamical
symmetry breaking associated with the memory of the quantum system.

\subsection*{Experimental setup}

\begin{figure}[hbpt!]
\centering
\includegraphics[width=0.66\linewidth]{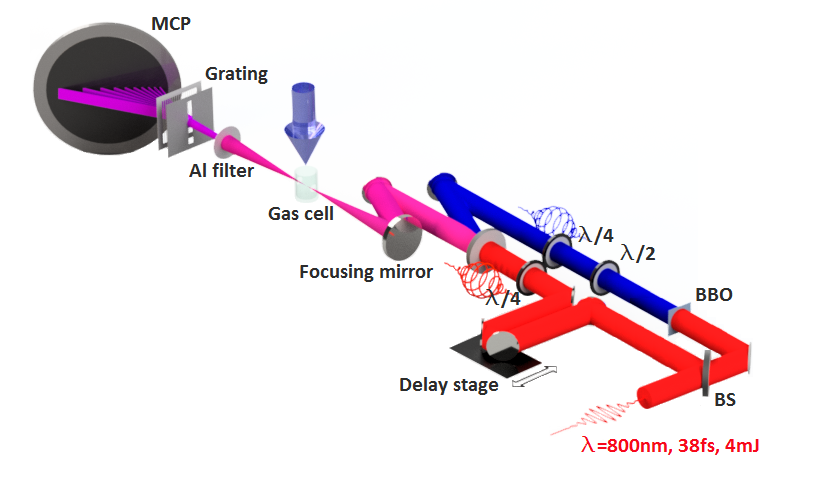}
\caption{\label{fig:setup} {\bf Experimenal layout.}
BBO: beta-barium-borate crystal for second harmonic generation, 
MCP: multi-channel plate based detector for XUV radiation,
BS: beam-splitting, partially transmitting mirror.}
\end{figure}

We have performed experiments in helium using a Ti:sapphire-based laser system with a single stage regenerative amplifier
producing 38~fs pulses with up to 4 mJ energy and a central wavelength of $\sim795$ nm at 1 kHz repetition rate.
The carrier-envelope phase (CEP) of the pulses was not locked. 
The laser beam was directed into the optical setup shown in Fig.~\ref{fig:setup}.

The original beam was split into two beams, with the possibility to use the splitting ratios of 50/50, 70/30 and 80/20.
The first beam was directed into a BBO crystal to generate the second harmonic (at $\sim 400$ nm) with the 
pulse energy up to 0.8 mJ.
The second beam, remaining at fundamental wavelength $795$ nm, 
had the pulse energy up to 1.0 mJ. 
We could also smoothly tune the energies
of the pulses and ratio between the `'red'' and the `'blue'' beams by changing the 
pump pulse energy in the amplifier.

Both beams passed through the corresponding achromatic broadband $\lambda/2$ and $\lambda/4$ waveplates,
where  their polarization was converted into nearly circular, with the ellipticity as high as $\varepsilon\simeq 0.95$.
The optical path of the fundamental beam was controlled using a rooftop mirror mounted on a translation
stage. The fundamental and the second harmonic beams were combined together in the 
collinear geometry and focused with a single Ag-mirror at f/100 into a 5-mm-long gas cell containing helium.

The waists of the fundamental and the second harmonic (red and blue)  beams were measured to 
be $w_0(\omega)\simeq 33$ $\mu$m and $w_0(2\omega)\simeq 27$
$\mu$m respectively, so that the maximum intensity could reach as high as  
$I_{\omega}\sim 9.0\times10^{14}$ $W/cm^2$ and $I_{2\omega}\sim 7.2\times10^{14}$ $W/cm^2$.
The pulses were focused approximately 2 mm before the target, minimizing 
the contribution of the Gouy phase to macroscopic effects and selecting short electron trajectories.  
The gas cell, placed inside the vacuum chamber, was initially sealed with a metal foil.
The foil was burned through by the laser beam at the start of the experiment.
The resulting cell opening  had the size $d=40~\mu$m similar to the spot size 
on the cell position, allowing us to keep the gas pressure inside  the cell 
constant at $\simeq  40$ mbar at the appropriate level of vacuum
(typically $P_{\rm rest}\approx10^{-4}$ mbar) inside the interaction chamber. 

After passing the 5 mm gas cell, the
driving `red' and `blue' beams were blocked by an $300$ nm thick aluminum foil.
The transmitted XUV radiation was directed towards the XUV spectrometer placed insight the 
vacuum chamber differentially separated from the interaction chamber. 
The XUV-spectromenter was based on the silicon nitride transmission nanograting
operating in the wavelength range of $10$ to $80$ nm \cite{korni} with a 
resolution of $0,25-0,13$ nm across the whole spectral range.
The generated and spectrally resolved XUV radiation was detected by a double-microchannel plate (MCP) 
with a phosphor screen and recorded by a fast CMOS camera (PointGrey). Radiation up to harmonic orders $\sim 50$ 
was observed.

\subsection*{Experimental results}

\begin{figure}[hbpt!]
\centering
\includegraphics[width=\linewidth]{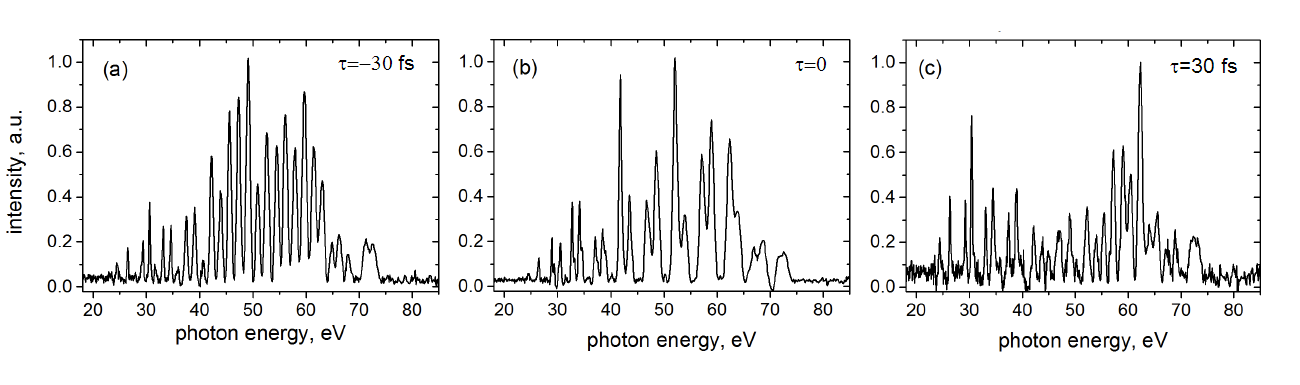}
\caption{\label{fig:delay} {\bf Experimenal spectra.}
Experimental XUV-spectra generated in the bi-circular field as a function of the 
$\omega-2\omega$ pulses time delay $\tau$ (negative delay corresponds to the blue pulse arriving first):
(a) $\tau$ =$- 30$ fs (close to 40 fsec cross-correlation length),
(b) $\tau$ =$0$ fs, 
(c) $\tau$ =$30$ fs.}
\end{figure}

Fig.~\ref{fig:delay} shows the observed XUV-spectra, for 
the bi-circular driving field and different red-blue time-delays. When the 
two driving pulses overlap (see Fig.~\ref{fig:delay} (b)), the harmonics with order $3N$ are suppressed.

We have measured and analysed the XUV spectra as a function of the 
time delay between the 800~nm and the 400~nm pulses.
The time zero of the perfect overlap between the two pulses
was determined via the cross-correlation between the two  beams in the BBO crystal, 
with the measured cross-correlation length  $\simeq 40$ fs.
The positive time delay means that the second harmonics arrives after the fundamental.

While the experimentally observed spectra are likely affected by 
the macroscopic propagation effects, we focus on the features that
must originate in the single-atom response: the dependence of
the forbidden harmonic lines on the delay and the relative intensities of
the two driving fields. Obviously, macroscopic propagation cannot lead
to the appearance of forbidden harmonics if they are not generated at the
single-atom level.

From this perspective, the main features in the 
experimental spectra in Fig.~\ref{fig:delay} are as follows. 
First, we see the appearance of the forbidden $3N$ harmonics when we increase the time-delay between the 
two pulses, while at $\tau$=0 these harmonics are very strongly suppressed. 
Second, the $3N$ harmonics are a lot more prominent for the negative time delay, i.e.
when the blue pulse arrives first. This experimentally observed feature was found to be 
robust with respect to varying the intensities of the two-color laser field.

\begin{figure}[hbpt!]
\centering
\includegraphics[width=\linewidth]{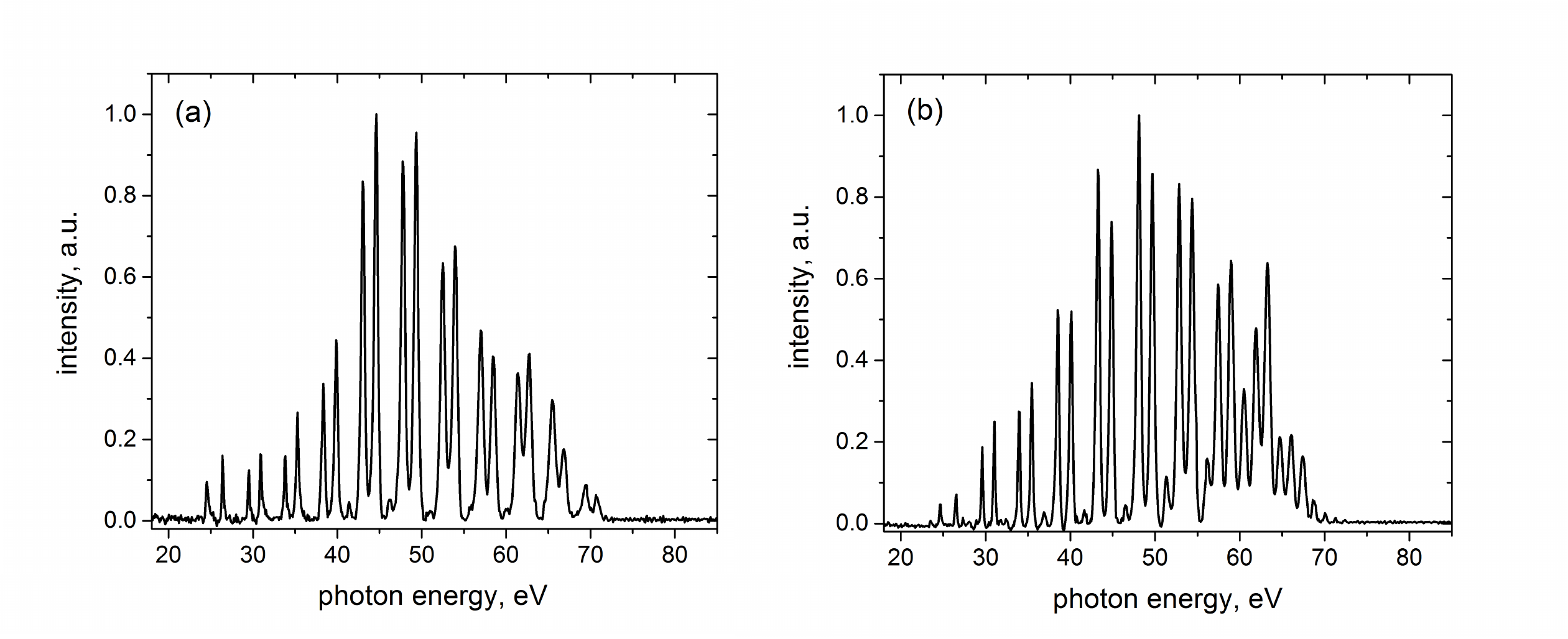}
\caption{\label{fig:inten} {\bf Experimenal spectra.}
Experimental XUV-spectra generated in the bi-circular field at $\tau$ =$0$ fs.
Panel (a): stronger red field, $I_{\omega}\sim 5 
\times10^{14}$ W/cm$^2$, $I_{2\omega}\sim 4 \times10^{14}$ W/cm$^2$);
Panel (b): stronger blue field, $I_{\omega}\sim 5 
\times10^{14}$ W/cm$^2$, $I_{2\omega}\sim 6 \times10^{14}$ W/cm$^2$)
}
\end{figure}

Fig. ~\ref{fig:inten} shows how changing the light intensity, 
especially the  ratio between the fundamental and the second harmonic,
affects the high harmonic spectrum. Apart from the clear effect of absorption in the lower
energy region and the expected trend that 
higher intensities leads to higher 
harmonic cut-offs,  we again bring the reader's attention to the 
forbidden $3N$ orders. 
We observe, that the forbidden $3N$ harmonics are effectively suppressed when the fundamental 
field is stronger than the second harmonic 
(Fig. ~\ref{fig:inten}a), or when the two intensities are 
close to each other (Fig.~\ref{fig:delay}b). 
When the intensity of the 400~nm pulse is higher than that of the 800~nm pulse, however, the 
forbidden harmonics become prominent (Fig. ~\ref{fig:inten}b).
These observations again seem to confirm the idea that frustrated tunneling is playing a prominent role.


\section*{Numerical Results}

We now turn to numerical simulations to reinforce
this physical idea and rule out propagation effects. The strong field approximation, which neglects the excited states, is not
adequate for the analysis, and we solve the time-dependent Schr\"odinger equation
for the Helium atom.

We used the code described in \cite{Patchkovskii2016}. To simulate the helium atom, we used the 3D single-active electron pseudo-potential given in \cite{Tong2005}. We have used a radial box of 600~a.u., with a total number of points $nr=1535$. We use a uniform grid, with 33 points (grid spacing of 0.14~a.u.) at the origin, followed by 34 points on a logarithmic grid, with a scaling parameter of 1.03, starting at 5~a.u., and finally 1468 points on a uniform grid with a spacing of 0.4~a.u. We placed a complex boundary absorber at the border of the radial box (starting at 470~a.u.), in order to avoid reflections. However, the box is sufficiently large to contain the full wave-function at the end of the pulse (we check that the total norm in the simulation volume is 1.0 at the end of the pulse). Therefore we can apply the iSURFC method \cite{Morales2016}. The maximum angular momenta included in the spherical harmonics expansion was $\ell_{\text{max}} = 70$. The time grid had a spacing of $dt = 0.04$~a.u. All the discretization parameters have been checked for convergence.

\begin{figure*}[hbpt!]
\centering
\includegraphics[width=\linewidth]{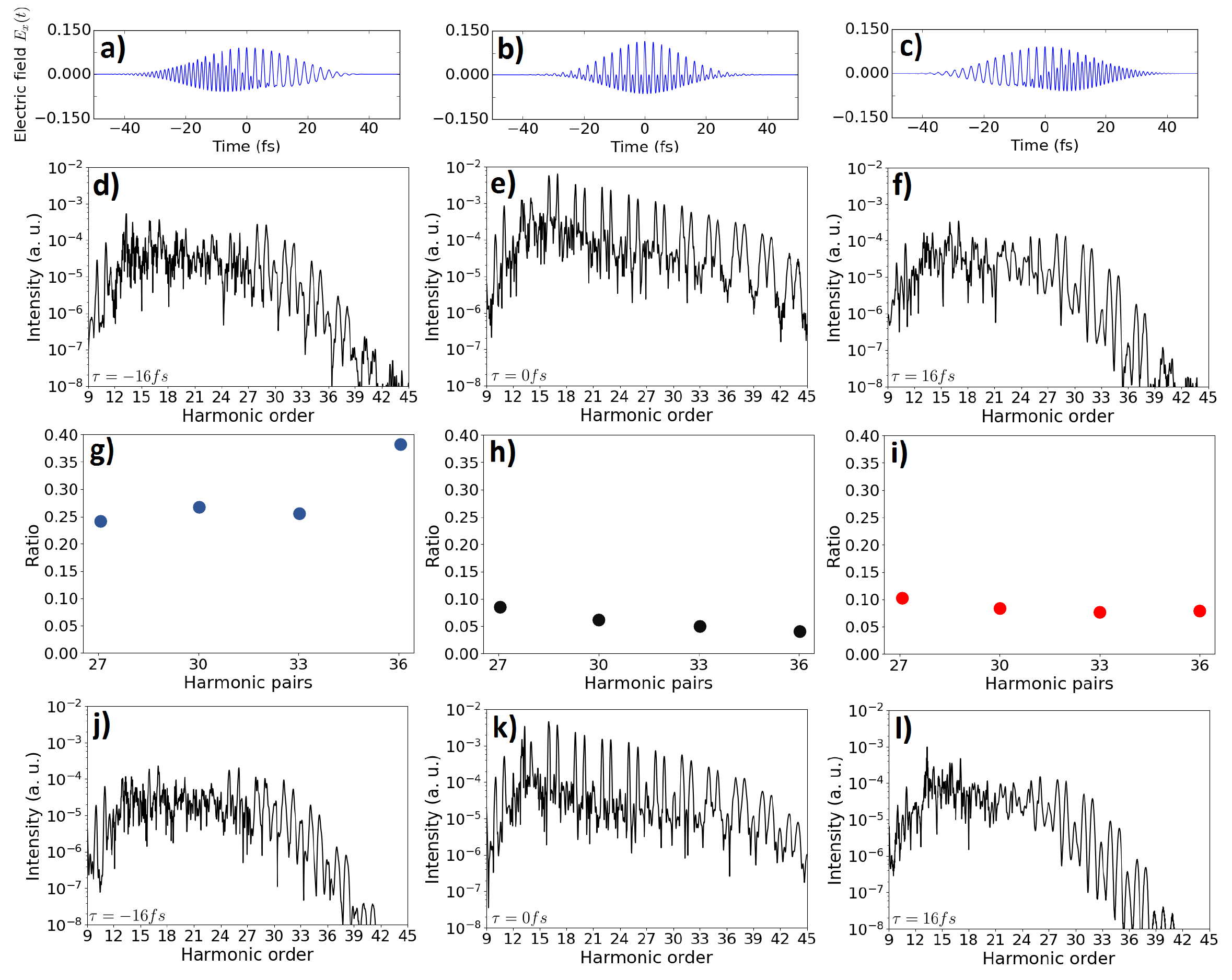}
\caption{\label{fig:timedelay} {\bf Theoretical HHG spectra as a function of the blue-red time delay.}
Top row: x-component of the bicircular field; second row: spectra for perfectly circular driving pulses;
third row: ratio of the forbidden 3N harmonic to its 3N+1 neighbor; fourth row: spectra for elliptical pulses, $\epsilon=0.9$.
Left column: the 400 nm  pulse comes first; middle
column: perfect overlap; right column: the 400 nm pulse is delayed.}
\end{figure*}

Fig.~\ref{fig:timedelay} shows our results obtained for 20~fs Gaussian pulses
with intensities $I_\omega^{\mathrm{th}}=I_{2\omega}^{\mathrm{th}}=0.12~$PW/cm$^2$,
corresponding to the peak fields of $F_\omega^{\mathrm{th}}=F_{2\omega}^{\mathrm{th}}=0.058$ a.u.,
with variable time delay of $\tau=-16,0,16$ fsec. The center of the fundamental pulse is
fixed at $t_{0,\omega}=0$, and $\tau=t_{0,2\omega}$ markes the
center of the second harmonic pulse. Positive $\tau$ means that  the second harmonic pulse comes later,
negative $\tau$ means that it comes earlier.
The top row shows the x-component of the total field, while the other rows
show the harmonic spectra and the ratio between the forbidden and permitted lines.
The spectra are presented for both perfectly circular
(panels d-f) and elliptic $\epsilon=0.9$ (panels j-l) fields, and the intensities of the two fields are equal.

As expected, the 3$N$ lines are suppressed for $\tau=0$ but become stronger as we increase
time delays between the two pulses. Crucially, the ratio $R(\tau)=S_{3N}(\tau)/S_{3N+1}(\tau)$
(panels g-i in Fig.~\ref{fig:timedelay})
is asymmetric as a function of $\tau$, strongly suggesting that the memory of the quantum
system is playing a role, i.e. the blue pulse excites the system and the delayed red pulse
probes the excitation. The effect is common for both perfectly circular and elliptic fields, with
the pulse ellipticity playing a secondary role in the effect.

\begin{figure}[hbpt!]
\centering
\includegraphics[width=\linewidth]{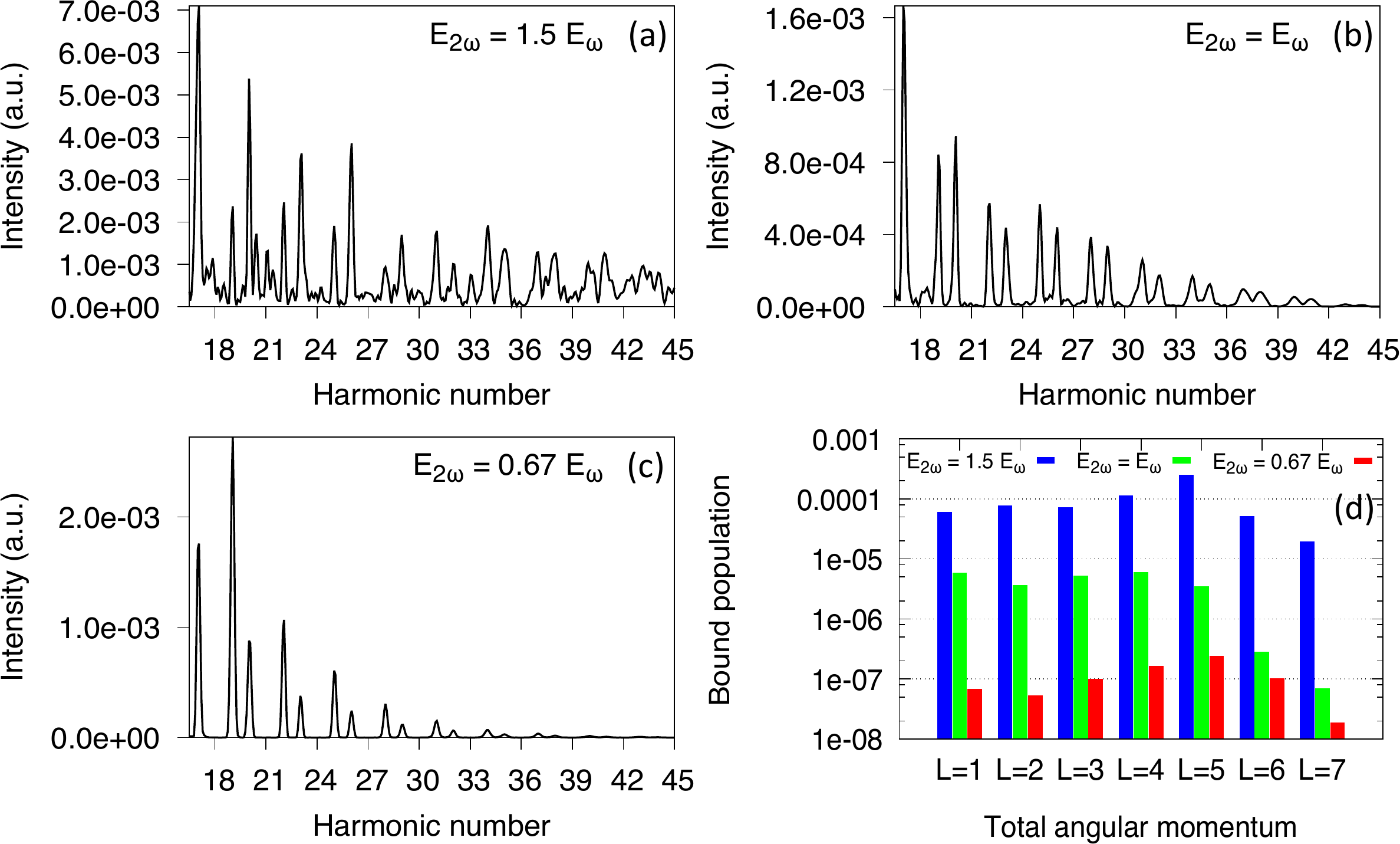}
\caption{\label{fig:bound} {\bf Influence of bound state population.}
High harmonic spectrum of helium generated by two counter-rotating $12$~fs long
fields of frequencies $1.55$~eV and $3.1$~eV, for three different ratios of the
field strengths:
(a) $F_{2\omega} = 3/2 F_{\omega}=0.085$,
(b) $F_{2\omega} = F_{\omega} = 0.057$, and
(c) $F_{2\omega} = 2/3 F_{\omega} =0.038$.
Peak intensities are the same for (a) and (c) and it is lower for (b), see text for details. 
Panel (d) shows the population of the bound states of the atom (in logarithmic scale)
after the interaction with the field, for the three intensity ratios.}
\end{figure}

We now focus on perfectly circular and perfectly overlapping pulses and show how
the forbidden harmonics arise even in these cases, thanks to the role of strongly driven
Rydberg states trapping population during strong field ionization. To this end,
we have performed theoretical simulations with perfectly circular, overlapping
pulses of 12~fs FWHM duration for
three different ratios of the field strengths, see Fig.~\ref{fig:bound}.
The ratio  $F_{2\omega}/F_{\omega}$ is varied from  $F_{2\omega}/F_{\omega}=2/3$ (c)
through $F_{2\omega}/F_{\omega}=1$ (b) to $F_{2\omega}/F_{\omega}=3/2$ (a).
The total intensity and, hence, the peak of the total electric field are kept constant
for (a) and (c), $I_{\max} = I_{2\omega} + I_{\omega} = 3.7 \times 10^{14}$~W/cm$^2$, and
is lowered for (b),  $I_{\max} = I_{2\omega} + I_{\omega} = 2.4 \times 10^{14}$~W/cm$^2$.

Substantial 3$N$ lines such as $H24, H27, H30$ and especially $H33$ appear when the second
harmonic is stronger than the fundamental  ($F_{2\omega} = (3/2)F_{\omega} =$ 0.085 a.u.)
and dominates ionization, see panel (a).
When the strength of the fundamental field field strengths are equal ($F_{2\omega} = F_{\omega} =$ 0.057 a.u.,),
panel (a), or when the fundamental is stronger ($F_{2\omega} = (2/3)F_{\omega} =$ 0.038 a.u.)
the 3$N$ harmonics are essentially absent.

To understand the reason behind this breaking of the symmetry, we projected the
wavefunction at the end of the pulse onto the bound
states of the atom. In Fig.~\ref{fig:bound}d we show the bound population of the first seven
excited states  (excluding the ground state), sorted by total angular momentum, at the end of the pulse.
When the blue field is stronger than the red field, irrespective of the maximum peak intensity,
the bound state population is dramatically higher.
In the energy domain (multiphoton) picture, this is the consequence of fewer high energy photons
needed to resonantly populate the higher lying states.
In the time-domain (tunneling) picture,  when the blue field dominates ionization,
the electron orbit is more likely to be trapped -- the frustrated tunnelling is more efficient
at 400 nm than at 800 nm.

\begin{figure}[hbpt!]
\centering
\includegraphics[width=\linewidth]{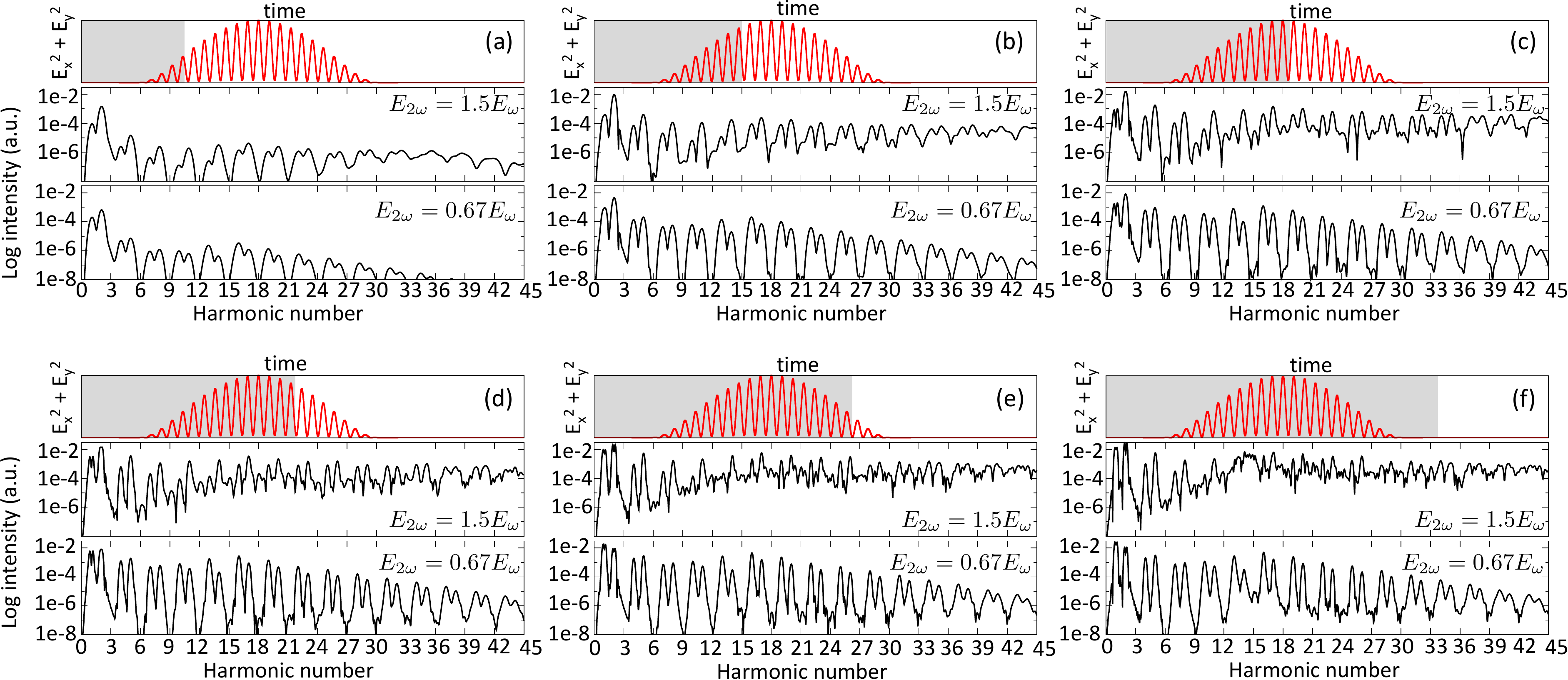}
\caption{\label{fig:movie} {\bf Build up of the spectrum.} Snapshots of 
the spectra in Fig.~\ref{fig:bound}a (central panel) and Fig.~\ref{fig:bound}c (bottom panel), 
at different times in the presence of the $12$~fs long laser field: 
(a) 6~fs before the peak of the field, (b) 2.42~fs before the peak of the field, 
(c) 0.6~fs after the peak of the field, (d) 3~fs after the peak of the field, 
(e) 6.65~fs after the peak of the field, and (f) 12.7~fs after the peak of the field.
The upper panel shows the maximum intensity of the field, $I_{max} = F_x^2+F_y^2$, which is 
always the same for the upper and lower spectra, as the red line, and the 
shaded area indicates the temporal window applied to the dipole to obtain the corresponding spectrum.}
\end{figure}

This picture is confirmed in Fig.~\ref{fig:movie}, where we show six snapshots of how the
spectrum in Fig.~\ref{fig:bound}a and \ref{fig:bound}c builds with time.
To do this, we apply a gradually increasing window function to the
time-dependent induced dipole $D(t)=\langle \Psi(t)|\hat d|\Psi(t)\rangle$. The
upper panel shows the length of this temporal window as the shaded area, along with the
total intensity $F_x^2 + F_y^2$ of the $\omega+2\omega$ laser field (red line).
The central and bottom panels show the corresponding spectrum for that energy window
for the cases when the 400 nm field is stronger or weaker than the 800 nm field, respectively.

Early on, Fig.~\ref{fig:movie}(a), the lower harmonics show only
symmetry-allowed harmonics.  For the higher harmonics, there is  not yet
enough time to provide a sequence of consecutive bursts
with sufficient energy,  which would interfere to yield clear
harmonic lines. The forbidden harmonics are
absent or very low in both spectra, indicating that the excited states are not sufficiently populated to
play any significant role. As we increase the temporal window (Fig.~\ref{fig:movie}b), the excited states
start to get populated when the blue field is stronger, and the electrons begin to get trapped in trajectories
orbiting around the ionic core. The forbidden harmonics start to emerge when the blue
field is stronger, but not when the red is stronger. The prohibited lines appear first at higher
harmonics. As we keep increasing the temporal
window (Fig.~\ref{fig:movie}c), lower harmonics start to show the forbidden lines.

More importantly, with increased time resolution
the forbidden lines such as H30, H33 or H36 in Fig.~\ref{fig:bound}c  start to
show a doublet structure. This is a characteristic feature of symmetry-forbidden lines, known
for single-color fields~\cite{ivanov1993coherent} and demonstrating the population of
more than one Floquet state during the laser pulse.
As time keeps increasing and the field
becomes stronger, the doublet lines start to appear also for lower harmonics
(see H18 and H21 in Fig.~\ref{fig:movie}d and H6 and H9 in Fig.~\ref{fig:movie}e).
Higher harmonics now start to show  a more complex structure,
suggesting that multiple Floquet states are being populated by the rapidly changing
field.
 While similar arguments are applicable for the case in which the red field is stronger,
 and indeed the forbidden harmonic lines are observable, the strengths of the signal is
 orders of magnitudes smaller due to the very small population
of the excited states in the first place.

This information can be partially accessed experimentally by time-delaying the driving pulses.
When the overlap between the two pulses is small, the spectrum should be
similar  to that in the case of the the short Fourier transform window, Fig.~\ref{fig:movie}a.
As the overlap of the pulses increases, the spectrum features will build up
as in Fig.~\ref{fig:movie}. With the blue pulse coming first, the
excited population will be higher, leading to more prominent forbidden lines
as discussed in the previous section.

%

\begin{figure}[hbpt!]
\centering
\includegraphics[width=0.5\linewidth]{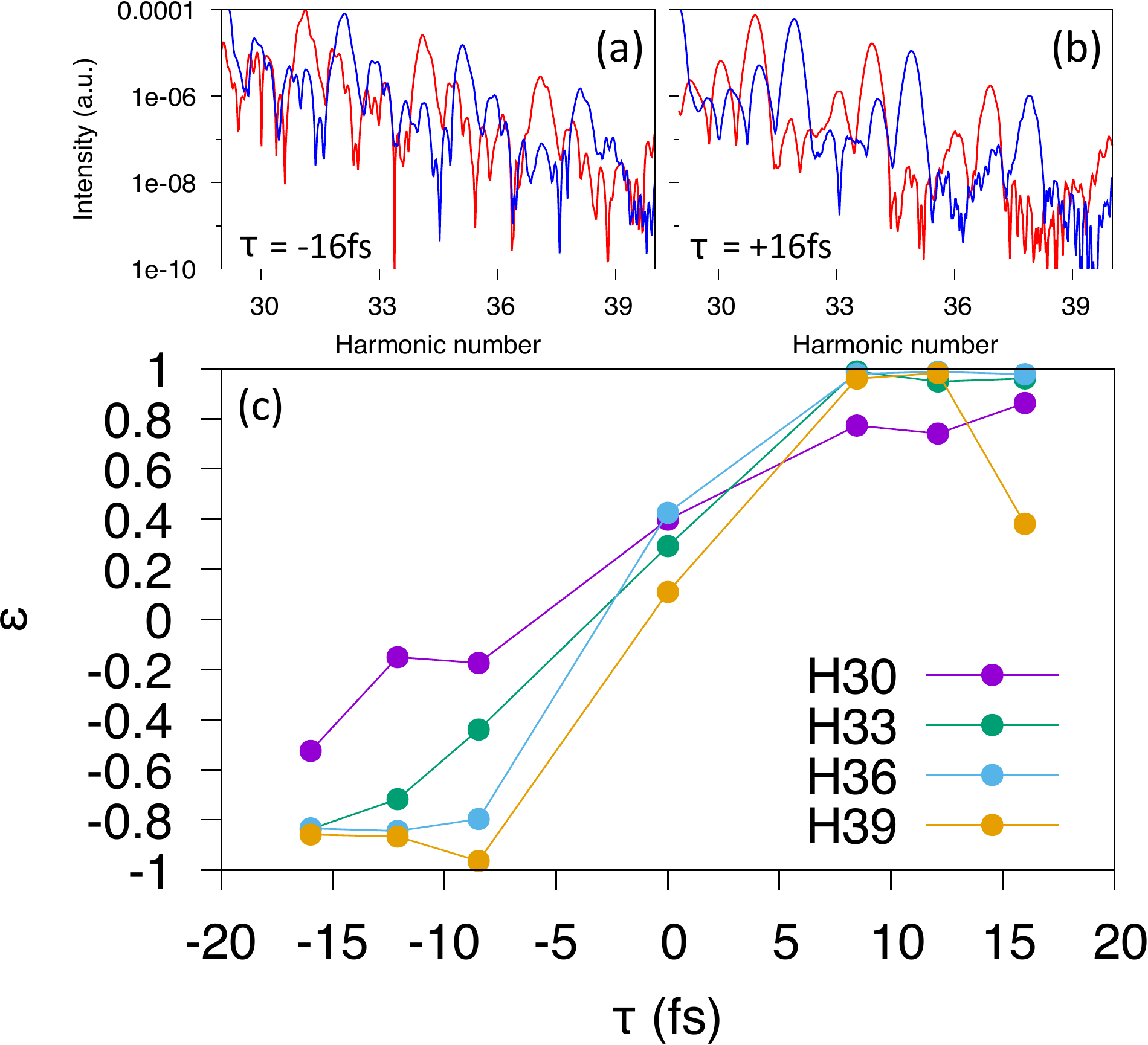}
\caption{\label{fig:ellip} {\bf Ellipticity of the forbidden harmonics.} Top panels: Close-up of the theoretical HHG spectrum of the four most clearly observed forbidden high harmonics in the spectrum (30,33,36,39), for a time delay in which the blue field comes 16~fs before (a) and after (b) the red field. The red line indicates the intensity component that co-rotates with the red field, while the blue line indicates the component that co-rotates with the blue field. In panel (c), the value of the ellipticity (as defined by Eq.~\ref{eq:ellip}) of each of the four forbidden harmonics, as a function of the time delay. The points indicate the values for which there is data, and the lines are used to guide the eye.}
\end{figure}

In conclusion of the theoretical analysis, we also point out that varying the time delay between the
two driving pulses allows us to control not only the strength of the
forbidden harmonics, but also their ellipticity. Here we define the ellipticity as
\begin{equation}\label{eq:ellip}
\epsilon = \frac{I_{\hat{e}_+}-I_{\hat{e}_-}}{I_{\hat{e}_+}+I_{\hat{e}_-}}.
\end{equation}
In Fig.~\ref{fig:ellip} we show how this value  changes as a function of the time delay for the four most
visible forbidden harmonics in the spectrum: 30, 33, 36 and 39. A clear trend is observed. The forbidden harmonics rotate preferentially with the field that comes first, thus providing a mechanism to coherently control their ellipticity.

Finally, we also observe strong blue-shift for higher harmonics when the blue light comes first. As we can see from the field profile in the top panel of Fig.~\ref{fig:timedelay}, in this case  the more intense bursts are
happen during the rising part of the field, leading to the blue shift of the harmonic
lines just like in the case of linearly polarized few-cycle drivers.
The opposite is true when red comes last.
Indeed, a high harmonic H37 experiences a strong blue-shift of $E_{H37}(\tau=-16~fs) - E_{H37} (\tau=16~fs) = 0.34$~eV. The situation seems to reverse for lower harmonics. A lower harmonic H25 slightly red-shifts its position by $E_{H25}(\tau=-16~fs) - E_{H25} (\tau=16~fs) = -0.12$~eV.  This provides a mechanism to fine tune the relative energy distance between harmonics.

\section*{Conclusion}

In conclusion, we have shown the different mechanisms by which the forbidden harmonic lines may appear in the high harmonic spectrum generated by bicircular fields. In contrast to the commonly accepted wisdom
that strong forbidden harmonics originate from slight ellipticity of the driving fields, we show that this is
not the case. For ellipticities $\epsilon\approx 0.95$ dynamical symmetry breaking is too weak to
be fully responsible for the strong forbidden lines.
We have demonstrated that population of Rydberg states breaks the dynamical symmetry and
leads to prominent signals at the forbidden harmonics.


The population of laser-driven Rydberg states is revealed by tracking the strength of the
forbidden harmonic lines via the time delay between the two driving pulses.
Increasing the intensity of the second harmonic field leads to stronger
trapping of the electrons in high Rydberg orbits. In time domain, this is
the consequence of the frustrated tunneling mechanism. In the frequency domain, they
can be seen as the Freeman resonances \cite{Gibson1992, Gibson1994}.
We have demonstrated that such dynamics are mapped on the  forbidden harmonic lines
that appear even for perfectly circular, long driving pulses.

We have temporally resolved these dynamics by applying a gradually increasing
window function to the Fourier transform of the induced dipole, demonstrating the build-up of the forbidden
harmonic lines as the excited states are populated on the rising edge of the laser pulse.
Finally, in analogy with the blue-shift observed in the high harmonic generation triggered by short linearly polarized drivers, we have predicted and experimentally confirmed a substantial blue-shift
of higher harmonics when the blue driver precedes the red driver.


\begin{thebibliography}{10}
\newcommand{\enquote}[1]{``#1''}

\bibitem{Krausz2009}
F.~Krausz and M.~Ivanov, \enquote{Attosecond physics,} Reviews of Modern
  Physics \textbf{81}, 163--234 (2009).

\bibitem{Sansone2010}
G.~Sansone, F.~Kelkensberg, J.~F. Perez-Torres, F.~Morales, M.~F. Kling,
  W.~Siu, O.~Ghafur, P.~Johnsson, M.~Swoboda, E.~Benedetti, F.~Ferrari,
  F.~Lepine, J.~L. Sanz-Vicario, S.~Zherebtsov, I.~Znakovskaya, A.~L/'Huillier,
  M.~Y. Ivanov, M.~Nisoli, F.~Martin, and M.~J.~J. Vrakking, \enquote{Electron
  localization following attosecond molecular photoionization,} Nature
  \textbf{465}, 763--766 (2010).

\bibitem{Gruson2016}
V.~Gruson, L.~Barreau, {\'A}.~Jim{\'e}nez-Galan, F.~Risoud, J.~Caillat,
  A.~Maquet, B.~Carr{\'e}, F.~Lepetit, J.~F. Hergott, T.~Ruchon, L.~Argenti,
  R.~Ta{\"\i}eb, F.~Mart{\'\i}n, and P.~Sali{\`e}res, \enquote{Attosecond
  dynamics through a fano resonance: Monitoring the birth of a photoelectron,}
  Science \textbf{354}, 734 (2016).

\bibitem{Kelkensberg2011}
F.~Kelkensberg, W.~Siu, J.~F. P{\'e}rez-Torres, F.~Morales, G.~Gademann,
  A.~Rouz{\'e}e, P.~Johnsson, M.~Lucchini, F.~Calegari, J.~L. Sanz-Vicario,
  F.~Mart{\'\i}n, and M.~J.~J. Vrakking, \enquote{Attosecond control in
  photoionization of hydrogen molecules,} Physical Review Letters \textbf{107},
  043002-- (2011).

\bibitem{Hassan2016}
M.~T. Hassan, T.~T. Luu, A.~Moulet, O.~Raskazovskaya, P.~Zhokhov, M.~Garg,
  N.~Karpowicz, A.~M. Zheltikov, V.~Pervak, F.~Krausz, and E.~Goulielmakis,
  \enquote{Optical attosecond pulses and tracking the nonlinear response of
  bound electrons,} Nature \textbf{530}, 66--70 (2016).

\bibitem{Garg2016}
M.~Garg, M.~Zhan, T.~T. Luu, H.~Lakhotia, T.~Klostermann, A.~Guggenmos, and
  E.~Goulielmakis, \enquote{Multi-petahertz electronic metrology,} Nature
  \textbf{538}, 359--363 (2016).

\bibitem{baker2006probing}
S.~Baker, J.~S. Robinson, C.~Haworth, H.~Teng, R.~Smith, C.~Chiril{\u{a}},
  M.~Lein, J.~Tisch, and J.~P. Marangos, \enquote{Probing proton dynamics in
  molecules on an attosecond time scale,} Science \textbf{312}, 424--427
  (2006).

\bibitem{smirnova2009high}
O.~Smirnova, Y.~Mairesse, S.~Patchkovskii, N.~Dudovich, D.~Villeneuve,
  P.~Corkum, and M.~Y. Ivanov, \enquote{High harmonic interferometry of
  multi-electron dynamics in molecules,} Nature \textbf{460}, 972--977 (2009).

\bibitem{haessler2010attosecond}
S.~Haessler, J.~Caillat, W.~Boutu, C.~Giovanetti-Teixeira, T.~Ruchon,
  T.~Auguste, Z.~Diveki, P.~Breger, A.~Maquet, B.~Carre, R.~Taieb, and
  P.~Salieres, \enquote{Attosecond imaging of molecular electronic
  wavepackets,} Nat Phys \textbf{6}, 200--206 (2010).

\bibitem{vozzi2011generalized}
C.~Vozzi, M.~Negro, F.~Calegari, G.~Sansone, M.~Nisoli, S.~De~Silvestri, and
  S.~Stagira, \enquote{Generalized molecular orbital tomography,} Nature
  Physics \textbf{7}, 822--826 (2011).

\bibitem{shafir2012resolving}
D.~Shafir, H.~Soifer, B.~D. Bruner, M.~Dagan, Y.~Mairesse, S.~Patchkovskii,
  M.~Y. Ivanov, O.~Smirnova, and N.~Dudovich, \enquote{Resolving the time when
  an electron exits a tunnelling barrier,} Nature \textbf{485}, 343--346
  (2012).

\bibitem{worner2010following}
H.~W{\"o}rner, J.~Bertrand, D.~Kartashov, P.~Corkum, and D.~Villeneuve,
  \enquote{Following a chemical reaction using high-harmonic interferometry,}
  Nature \textbf{466}, 604--607 (2010).

\bibitem{ferre2016two}
A.~Ferr{\'e}, H.~Soifer, O.~Pedatzur, C.~Bourassin-Bouchet, B.~D. Bruner,
  R.~Canonge, F.~Catoire, D.~Descamps, B.~Fabre, E.~M{\'e}vel, S.~Petit,
  N.~Dudovich, and Y.~Mairesse, \enquote{Two-dimensional frequency resolved
  optomolecular gating of high-order harmonic generation,} Physical Review
  Letters \textbf{116}, 053002-- (2016).

\bibitem{Morales2012}
F.~Morales, I.~Barth, V.~Serbinenko, S.~Patchkovskii, and O.~Smirnova,
  \enquote{Shaping polarization of attosecond pulses via laser control of
  electron and hole dynamics,} Journal of Modern Optics \textbf{59}, 1303--1311
  (2012).

\bibitem{pedatzur2015attosecond}
O.~Pedatzur, G.~Orenstein, V.~Serbinenko, H.~Soifer, B.~D. Bruner, A.~J. Uzan,
  D.~S. Brambila, A.~G. Harvey, L.~Torlina, F.~Morales, O.~Smirnova, and
  N.~Dudovich, \enquote{Attosecond tunnelling interferometry,} Nat Phys
  \textbf{11}, 815--819 (2015).

\bibitem{bruner2015multidimensional}
B.~D. Bruner, H.~Soifer, D.~Shafir, V.~Serbinenko, O.~Smirnova, and
  N.~Dudovich, \enquote{Multidimensional high harmonic spectroscopy,} Journal
  of Physics B: Atomic, Molecular and Optical Physics \textbf{48}, 174006
  (2015).

\bibitem{bruner2016multidimensional}
B.~D. Bruner, Z.~Masin, M.~Negro, F.~Morales, D.~Brambila, M.~Devetta,
  D.~Facciala, A.~G. Harvey, M.~Ivanov, Y.~Mairesse, S.~Patchkovskii,
  V.~Serbinenko, H.~Soifer, S.~Stagira, C.~Vozzi, N.~Dudovich, and O.~Smirnova,
  \enquote{Multidimensional high harmonic spectroscopy of polyatomic molecules:
  detecting sub-cycle laser-driven hole dynamics upon ionization in strong
  mid-ir laser fields,} Faraday Discuss. \textbf{194}, 369--405 (2016).

\bibitem{serbinenko2013multidimensional}
V.~Serbinenko and O.~Smirnova, \enquote{Multidimensional high harmonic
  spectroscopy: a semi-classical perspective on measuring multielectron
  rearrangement upon ionization,} Journal of Physics B: Atomic, Molecular and
  Optical Physics \textbf{46}, 171001 (2013).

\bibitem{Smirnova2015}
O.~Smirnova, Y.~Mairesse, and S.~Patchkovskii, \enquote{Opportunities for
  chiral discrimination using high harmonic generation in tailored laser
  fields,} Journal of Physics B: Atomic, Molecular and Optical Physics
  \textbf{48}, 234005 (2015).

\bibitem{kraus2015measurement}
P.~M. Kraus, B.~Mignolet, D.~Baykusheva, A.~Rupenyan, L.~Horn{\'y}, E.~F.
  Penka, G.~Grassi, O.~I. Tolstikhin, J.~Schneider, F.~Jensen, L.~B. Madsen,
  A.~D. Bandrauk, F.~Remacle, and H.~J. W{\"o}rner, \enquote{Measurement and
  laser control of attosecond charge migration in ionized iodoacetylene,}
  Science \textbf{350}, 790--795 (2015).

\bibitem{worner2011conical}
H.~J. W{\"o}rner, J.~B. Bertrand, B.~Fabre, J.~Higuet, H.~Ruf, A.~Dubrouil,
  S.~Patchkovskii, M.~Spanner, Y.~Mairesse, V.~Blanchet, E.~M{\'e}vel,
  E.~Constant, P.~B. Corkum, and D.~M. Villeneuve, \enquote{Conical
  intersection dynamics in no{$<$}sub{$>$}2{$<$}/sub{$>$} probed by homodyne
  high-harmonic spectroscopy,} Science \textbf{334}, 208 (2011).

\bibitem{ghimire2011observation}
S.~Ghimire, A.~D. DiChiara, E.~Sistrunk, P.~Agostini, L.~F. DiMauro, and D.~A.
  Reis, \enquote{Observation of high-order harmonic generation in a bulk
  crystal,} Nature physics \textbf{7}, 138--141 (2011).

\bibitem{schiffrin2013optical}
A.~Schiffrin, T.~Paasch-Colberg, N.~Karpowicz, V.~Apalkov, D.~Gerster,
  S.~Muhlbrandt, M.~Korbman, J.~Reichert, M.~Schultze, S.~Holzner, J.~V. Barth,
  R.~Kienberger, R.~Ernstorfer, V.~S. Yakovlev, M.~I. Stockman, and F.~Krausz,
  \enquote{Optical-field-induced current in dielectrics,} Nature \textbf{493},
  70--74 (2013).

\bibitem{ivanov2013opportunities}
M.~Ivanov and O.~Smirnova, \enquote{Opportunities for sub-laser-cycle
  spectroscopy in condensed phase,} Chemical Physics \textbf{414}, 3--9 (2013).

\bibitem{vampa2014theoretical}
G.~Vampa, C.~McDonald, G.~Orlando, D.~Klug, P.~Corkum, and T.~Brabec,
  \enquote{Theoretical analysis of high-harmonic generation in solids,}
  Physical review letters \textbf{113}, 073901 (2014).

\bibitem{vampa2015linking}
G.~Vampa, T.~Hammond, N.~Thir{\'e}, B.~Schmidt, F.~L{\'e}gar{\'e}, C.~McDonald,
  T.~Brabec, and P.~Corkum, \enquote{Linking high harmonics from gases and
  solids,} Nature \textbf{522}, 462--464 (2015).

\bibitem{langer2016lightwave}
F.~Langer, M.~Hohenleutner, C.~P. Schmid, C.~Poellmann, P.~Nagler, T.~Korn,
  C.~Sch{\"u}ller, M.~S. Sherwin, U.~Huttner, J.~T. Steiner, S.~W. Koch,
  M.~Kira, and R.~Huber, \enquote{Lightwave-driven quasiparticle collisions on
  a subcycle timescale,} Nature \textbf{533}, 225--229 (2016).

\bibitem{Eichmann1995}
H.~Eichmann, A.~Egbert, S.~Nolte, C.~Momma, B.~Wellegehausen, W.~Becker,
  S.~Long, and J.~K. McIver, \enquote{Polarization-dependent high-order
  two-color mixing,} Physical Review A \textbf{51}, R3414--R3417 (1995).

\bibitem{Milosevic2000}
D.~B. Milo{\v s}evi{\'c}, W.~Becker, and R.~Kopold, \enquote{Generation of
  circularly polarized high-order harmonics by two-color coplanar field
  mixing,} Physical Review A \textbf{61}, 063403-- (2000).

\bibitem{fleischer2014spin}
A.~Fleischer, O.~Kfir, T.~Diskin, P.~Sidorenko, and O.~Cohen, \enquote{Spin
  angular momentum and tunable polarization in high-harmonic generation,}
  Nature Photonics \textbf{8}, 543--549 (2014).

\bibitem{Chen2016}
C.~Chen, Z.~Tao, C.~Hern{\'a}ndez-Garc{\'\i}a, P.~Matyba, A.~Carr, R.~Knut,
  O.~Kfir, D.~Zusin, C.~Gentry, P.~Grychtol, O.~Cohen, L.~Plaja, A.~Becker,
  A.~Jaron-Becker, H.~Kapteyn, and M.~Murnane, \enquote{Tomographic
  reconstruction of circularly polarized high-harmonic fields: 3d attosecond
  metrology,} Science Advances \textbf{2} (2016).

\bibitem{Hickstein2015}
D.~D. Hickstein, F.~J. Dollar, P.~Grychtol, J.~L. Ellis, R.~Knut,
  C.~Hern{\'a}ndez-Garc{\'\i}a, D.~Zusin, C.~Gentry, J.~M. Shaw, T.~Fan, K.~M.
  Dorney, A.~Becker, A.~Jaro{\'n}-Becker, H.~C. Kapteyn, M.~M. Murnane, and
  C.~G. Durfee, \enquote{Non-collinear generation of angularly isolated
  circularly polarized high harmonics,} Nat Photon \textbf{9}, 743--750 (2015).

\bibitem{Kfir2015}
O.~Kfir, P.~Grychtol, E.~Turgut, R.~Knut, D.~Zusin, D.~Popmintchev,
  T.~Popmintchev, H.~Nembach, J.~M. Shaw, A.~Fleischer, H.~Kapteyn, M.~Murnane,
  and O.~Cohen, \enquote{Generation of bright phase-matched
  circularly-polarized extreme ultraviolet high harmonics,} Nat Photon
  \textbf{9}, 99--105 (2015).

\bibitem{Baykusheva2016}
D.~Baykusheva, M.~S. Ahsan, N.~Lin, and H.~J. W{\"o}rner, \enquote{Bicircular
  high-harmonic spectroscopy reveals dynamical symmetries of atoms and
  molecules,} Physical Review Letters \textbf{116}, 123001-- (2016).

\bibitem{bandrauk2016circularly}
A.~D. Bandrauk, F.~Mauger, and K.-J. Yuan, \enquote{Circularly polarized
  harmonic generation by intense bicircular laser pulses: electron recollision
  dynamics and frequency dependent helicity,} Journal of Physics B: Atomic,
  Molecular and Optical Physics \textbf{49}, 23LT01 (2016).

\bibitem{mauger2016circularly}
F.~Mauger, A.~Bandrauk, and T.~Uzer, \enquote{Circularly polarized molecular
  high harmonic generation using a bicircular laser,} Journal of Physics B:
  Atomic, Molecular and Optical Physics \textbf{49}, 10LT01 (2016).

\bibitem{odvzak2016atomic}
S.~Od{\v{z}}ak, E.~Hasovi{\'c}, W.~Becker, and D.~Milo{\v{s}}evi{\'c},
  \enquote{Atomic processes in bicircular fields,} Journal of Modern Optics pp.
  1--10 (2016).

\bibitem{odvzak2016high}
S.~Od{\v{z}}ak, E.~Hasovi{\'c}, and D.~B. Milo{\v{s}}evi{\'c},
  \enquote{High-order harmonic generation in polyatomic molecules induced by a
  bicircular laser field,} Physical Review A \textbf{94}, 033419 (2016).

\bibitem{Fan2015}
T.~Fan, P.~Grychtol, R.~Knut, C.~Hern{\'a}ndez-Garc{\'\i}a, D.~D. Hickstein,
  D.~Zusin, C.~Gentry, F.~J. Dollar, C.~A. Mancuso, C.~W. Hogle, O.~Kfir,
  D.~Legut, K.~Carva, J.~L. Ellis, K.~M. Dorney, C.~Chen, O.~G. Shpyrko, E.~E.
  Fullerton, O.~Cohen, P.~M. Oppeneer, D.~B. Milo{\v s}evi{\'c}, A.~Becker,
  A.~A. Jaro{\'n}-Becker, T.~Popmintchev, M.~M. Murnane, and H.~C. Kapteyn,
  \enquote{Bright circularly polarized soft x-ray high harmonics for x-ray
  magnetic circular dichroism,} Proceedings of the National Academy of Sciences
  \textbf{112}, 14206--14211 (2015).

\bibitem{Milosevic2015}
D.~B. Milo{\v s}evi{\'c}, \enquote{Generation of elliptically polarized
  attosecond pulse trains,} Optics Letters \textbf{40}, 2381--2384 (2015).

\bibitem{Medisauskas2015}
L.~Medi{\v s}auskas, J.~Wragg, H.~van~der Hart, and M.~Y. Ivanov,
  \enquote{Generating isolated elliptically polarized attosecond pulses using
  bichromatic counterrotating circularly polarized laser fields,} Physical
  Review Letters \textbf{115}, 153001-- (2015).

\bibitem{Zhang2017}
X.~Zhang, X.~Zhu, X.~Liu, D.~Wang, Q.~Zhang, P.~Lan, and P.~Lu,
  \enquote{Ellipticity-tunable attosecond xuv pulse generation with a rotating
  bichromatic circularly polarized laser field,} Optics Letters \textbf{42},
  1027--1030 (2017).

\bibitem{Cireasa2015}
R.~Cireasa, A.~E. Boguslavskiy, B.~Pons, M.~C.~H. Wong, D.~Descamps, S.~Petit,
  H.~Ruf, N.~Thire, A.~Ferre, J.~Suarez, J.~Higuet, B.~E. Schmidt, A.~F.
  Alharbi, F.~Legare, V.~Blanchet, B.~Fabre, S.~Patchkovskii, O.~Smirnova,
  Y.~Mairesse, and V.~R. Bhardwaj, \enquote{Probing molecular chirality on a
  sub-femtosecond timescale,} Nat Phys \textbf{11}, 654--658 (2015).

\bibitem{Cavalieri2007}
A.~L. Cavalieri, N.~Muller, T.~Uphues, V.~S. Yakovlev, A.~Baltuska, B.~Horvath,
  B.~Schmidt, L.~Blumel, R.~Holzwarth, S.~Hendel, M.~Drescher, U.~Kleineberg,
  P.~M. Echenique, R.~Kienberger, F.~Krausz, and U.~Heinzmann,
  \enquote{Attosecond spectroscopy in condensed matter,} Nature \textbf{449},
  1029--1032 (2007).

\bibitem{Boeglin2010}
C.~Boeglin, E.~Beaurepaire, V.~Halte, V.~Lopez-Flores, C.~Stamm, N.~Pontius,
  H.~A. Durr, and J.~Y. Bigot, \enquote{Distinguishing the ultrafast dynamics
  of spin and orbital moments in solids,} Nature \textbf{465}, 458--461 (2010).

\bibitem{Graves2013}
C.~E. Graves, A.~H. Reid, T.~Wang, B.~Wu, S.~de~Jong, K.~Vahaplar, I.~Radu,
  D.~P. Bernstein, M.~Messerschmidt, L.~M{\"u}ller, R.~Coffee, M.~Bionta, S.~W.
  Epp, R.~Hartmann, N.~Kimmel, G.~Hauser, A.~Hartmann, P.~Holl, H.~Gorke, J.~H.
  Mentink, A.~Tsukamoto, A.~Fognini, J.~J. Turner, W.~F. Schlotter, D.~Rolles,
  H.~Soltau, L.~Str{\"u}der, Y.~Acremann, A.~V. Kimel, A.~Kirilyuk, T.~Rasing,
  J.~St{\"o}hr, A.~O. Scherz, and H.~A. D{\"u}rr, \enquote{Nanoscale spin
  reversal by non-local angular momentum transfer following ultrafast laser
  excitation in ferrimagnetic gdfeco,} Nat Mater \textbf{12}, 293--298 (2013).

\bibitem{dudovich2006measuring}
N.~Dudovich, O.~Smirnova, J.~Levesque, Y.~Mairesse, M.~Y. Ivanov,
  D.~Villeneuve, and P.~B. Corkum, \enquote{Measuring and controlling the birth
  of attosecond xuv pulses,} Nature physics \textbf{2}, 781--786 (2006).

\bibitem{mansten2008spectral}
E.~Mansten, J.~M. Dahlstr{\"o}m, P.~Johnsson, M.~Swoboda, A.~L'Huillier, and
  J.~Mauritsson, \enquote{Spectral shaping of attosecond pulses using
  two-colour laser fields,} New Journal of Physics \textbf{10}, 083041 (2008).

\bibitem{mauritsson2009sub}
J.~Mauritsson, J.~Dahlstr{\"o}m, E.~Mansten, and T.~Fordell, \enquote{Sub-cycle
  control of attosecond pulse generation using two-colour laser fields,}
  Journal of Physics B: Atomic, Molecular and Optical Physics \textbf{42},
  134003 (2009).

\bibitem{he2010interference}
X.~He, J.~Dahlstr{\"o}m, R.~Rakowski, C.~Heyl, A.~Persson, J.~Mauritsson, and
  A.~L'Huillier, \enquote{Interference effects in two-color high-order harmonic
  generation,} Physical Review A \textbf{82}, 033410 (2010).

\bibitem{dahlstrom2011quantum}
J.~Dahlstr{\"o}m, A.~L'Huillier, and J.~Mauritsson, \enquote{Quantum mechanical
  approach to probing the birth of attosecond pulses using a two-colour field,}
  Journal of Physics B: Atomic, Molecular and Optical Physics \textbf{44},
  095602 (2011).

\bibitem{brugnera2011trajectory}
L.~Brugnera, D.~J. Hoffmann, T.~Siegel, F.~Frank, A.~Za{\"\i}r, J.~W. Tisch,
  and J.~P. Marangos, \enquote{Trajectory selection in high harmonic generation
  by controlling the phase between orthogonal two-color fields,} Physical
  review letters \textbf{107}, 153902 (2011).

\bibitem{ganeev2012experimental}
R.~Ganeev, V.~Strelkov, C.~Hutchison, A.~Za{\"\i}r, D.~Kilbane, M.~Khokhlova,
  and J.~Marangos, \enquote{Experimental and theoretical studies of
  two-color-pump resonance-induced enhancement of odd and even harmonics from a
  tin plasma,} Physical Review A \textbf{85}, 023832 (2012).

\bibitem{ivanov1993coherent}
M.~Y. Ivanov, P.~Corkum, and P.~Dietrich, \enquote{Coherent control and
  collapse of symmetry in a two-level system in an intense laser field,} Laser
  Physics \textbf{3}, 375--380 (1993).

\bibitem{smirnova2007anatomy}
O.~Smirnova, M.~Spanner, and M.~Ivanov, \enquote{Anatomy of strong field
  ionization ii: to dress or not to dress?} Journal of Modern Optics
  \textbf{54}, 1019--1038 (2007).

\bibitem{morales2014high}
F.~Morales, P.~Rivi{\`e}re, M.~Richter, A.~Gubaydullin, M.~Ivanov, O.~Smirnova,
  and F.~Mart{\'\i}n, \enquote{High harmonic spectroscopy of electron
  localization in the hydrogen molecular ion,} Journal of Physics B: Atomic,
  Molecular and Optical Physics \textbf{47}, 204015 (2014).

\bibitem{bian2014probing}
X.-B. Bian and A.~D. Bandrauk, \enquote{Probing nuclear motion by frequency
  modulation of molecular high-order harmonic generation,} Physical review
  letters \textbf{113}, 193901 (2014).

\bibitem{Silva2016}
R.~E.~F. Silva, P.~Rivi{\`e}re, F.~Morales, O.~Smirnova, M.~Ivanov, and
  F.~Mart{\'\i}n, \enquote{Even harmonic generation in isotropic media of
  dissociating homonuclear molecules,} Scientific Reports \textbf{6}, 32653
  (2016).

\bibitem{yudin2001physics}
G.~L. Yudin and M.~Y. Ivanov, \enquote{Physics of correlated double ionization
  of atoms in intense laser fields: Quasistatic tunneling limit,} Physical
  Review A \textbf{63}, 033404 (2001).

\bibitem{nubbemeyer2008strong}
T.~Nubbemeyer, K.~Gorling, A.~Saenz, U.~Eichmann, and W.~Sandner,
  \enquote{Strong-field tunneling without ionization,} Physical review letters
  \textbf{101}, 233001 (2008).

\bibitem{eichmann2009acceleration}
U.~Eichmann, T.~Nubbemeyer, H.~Rottke, and W.~Sandner, \enquote{Acceleration of
  neutral atoms in strong short-pulse laser fields,} Nature \textbf{461},
  1261--1264 (2009).

\bibitem{von2013frustrated}
A.~von Veltheim, B.~Manschwetus, W.~Quan, B.~Borchers, G.~Steinmeyer,
  H.~Rottke, and W.~Sandner, \enquote{Frustrated tunnel ionization of noble gas
  dimers with rydberg-electron shakeoff by electron charge oscillation,}
  Physical review letters \textbf{110}, 023001 (2013).

\bibitem{eichmann2013observing}
U.~Eichmann, A.~Saenz, S.~Eilzer, T.~Nubbemeyer, and W.~Sandner,
  \enquote{Observing rydberg atoms to survive intense laser fields,} Physical
  review letters \textbf{110}, 203002 (2013).

\bibitem{zimmermann2017unified}
H.~Zimmermann, S.~Patchkovskii, M.~Ivanov, and U.~Eichmann, \enquote{Unified
  time and frequency picture of ultrafast atomic excitation in strong laser
  fields,} Physical Review Letters \textbf{118}, 013003 (2017).

\bibitem{richter2013role}
M.~Richter, S.~Patchkovskii, F.~Morales, O.~Smirnova, and M.~Ivanov,
  \enquote{The role of the kramers--henneberger atom in the higher-order kerr
  effect,} New Journal of Physics \textbf{15}, 083012 (2013).

\bibitem{popov2003strong}
A.~Popov, O.~Tikhonova, and E.~Volkova, \enquote{Strong-field atomic
  stabilization: numerical simulation and analytical modelling,} Journal of
  Physics B: Atomic, Molecular and Optical Physics \textbf{36}, R125 (2003).

\bibitem{popov2011different}
A.~Popov, O.~Tikhonova, and E.~Volkova, \enquote{Different regimes of
  strong-field dynamics of atoms in intense low-frequency laser pulses,}
  Journal of Modern Optics \textbf{58}, 1195--1205 (2011).

\bibitem{fedorov2012interference}
M.~V. Fedorov, N.~P. Poluektov, A.~M. Popov, O.~V. Tikhonova, V.~Y. Kharin, and
  E.~A. Volkova, \enquote{Interference stabilization revisited,} IEEE Journal
  of Selected Topics in Quantum Electronics \textbf{18}, 42--53 (2012).

\bibitem{morales2011imaging}
F.~Morales, M.~Richter, S.~Patchkovskii, and O.~Smirnova, \enquote{Imaging the
  kramers--henneberger atom,} Proceedings of the National Academy of Sciences
  \textbf{108}, 16906--16911 (2011).

\bibitem{bredtmann2016:xuv}
T.~Bredtmann, S.~Chelkowski, A.~D. Bandrauk, and M.~Ivanov, \enquote{Xuv lasing
  during strong-field-assisted transient absorption in molecules,} Physical
  Review A \textbf{93}, 021402 (2016).

\bibitem{Yudin2001}
G.~L. Yudin and M.~Y. Ivanov, \enquote{Physics of correlated double ionization
  of atoms in intense laser fields: Quasistatic tunneling limit,} Physical
  Review A \textbf{63}, 033404-- (2001).

\bibitem{manschwetus2009strong}
B.~Manschwetus, T.~Nubbemeyer, K.~Gorling, G.~Steinmeyer, U.~Eichmann,
  H.~Rottke, and W.~Sandner, \enquote{Strong laser field fragmentation of h 2:
  Coulomb explosion without double ionization,} Physical review letters
  \textbf{102}, 113002 (2009).

\bibitem{agostini2004physics}
P.~Agostini and L.~F. DiMauro, \enquote{The physics of attosecond light
  pulses,} Reports on progress in physics \textbf{67}, 813 (2004).

\bibitem{Ishikawa2003}
K.~Ishikawa, \enquote{Photoemission and ionization of
  {\$}{\{}{$\backslash$}mathrm{\{}h{\}}{$\backslash$}mathrm{\{}e{\}}{\}}\^{}{\{}+{\}}{\$}
  under simultaneous irradiation of fundamental laser and high-order harmonic
  pulses,} Physical Review Letters \textbf{91}, 043002-- (2003).

\bibitem{korni}
O.~Kornilov, R.~Wilcox, and O.~Gessner, \enquote{Nanograting-based compact
  vacuum ultraviolet spectrometer and beam profile for in situ characterization
  of high-order harmonic generation light sources,} Review of Scientific
  Instruments \textbf{81}, 063109--063112 (2010).

\bibitem{Patchkovskii2016}
S.~Patchkovskii and H.~G. Muller, \enquote{Simple, accurate, and efficient
  implementation of 1-electron atomic time-dependent schr{\"o}dinger equation
  in spherical coordinates,} Computer Physics Communications \textbf{199},
  153--169 (2016).

\bibitem{Tong2005}
X.~M. Tong and C.~D. Lin, \enquote{Empirical formula for static field
  ionization rates of atoms and molecules by lasers in the barrier-suppression
  regime,} Journal of Physics B: Atomic, Molecular and Optical Physics
  \textbf{38}, 2593 (2005).

\bibitem{Morales2016}
F.~Morales, T.~Bredtmann, and S.~Patchkovskii, \enquote{isurf: a family of
  infinite-time surface flux methods,} Journal of Physics B: Atomic, Molecular
  and Optical Physics \textbf{49}, 245001 (2016).

\bibitem{Gibson1992}
G.~N. Gibson, R.~R. Freeman, and T.~J. McIlrath, \enquote{Verification of the
  dominant role of resonant enhancement in short-pulse multiphoton ionization,}
  Physical Review Letters \textbf{69}, 1904--1907 (1992).

\bibitem{Gibson1994}
G.~N. Gibson, R.~R. Freeman, T.~J. McIlrath, and H.~G. Muller,
  \enquote{Excitation and ionization dynamics in short-pulse multiphoton
  ionization,} Physical Review A \textbf{49}, 3870--3874 (1994).

\end{thebibliography}

\end{document}